\documentclass{egpubl}
\usepackage{eurovis2019}

\SpecialIssuePaper         % uncomment for final version of Computer Graphics Forum, special issue
 \electronicVersion % can be used both for the printed and electronic version

\ifpdf \usepackage[pdftex]{graphicx} \pdfcompresslevel=9
\else \usepackage[dvips]{graphicx} \fi

\PrintedOrElectronic

\usepackage{t1enc,dfadobe}

\usepackage{multirow}
\usepackage{egweblnk}
\usepackage{amsmath}
\usepackage{amssymb}
\usepackage{cite}
\usepackage{verbatim}
\usepackage[T1]{fontenc}

\newcommand{\hone}{H~\textsc{i}}
\newcommand{\cfour}{C~\textsc{iv}}

\newcommand{\osix}{O~\textsc{vi}}

\newcommand{\rvir}{$r_{vir}$}
\newcommand{\lya}{Ly$\alpha$}

\usepackage{listings} %code highlighter
\usepackage{color} %use color
\definecolor{mygreen}{rgb}{0,0.6,0}
\definecolor{mygray}{rgb}{0.5,0.5,0.5}
\definecolor{mymauve}{rgb}{0.58,0,0.82}
 
\definecolor{darkgray}{rgb}{.4,.4,.4}
\definecolor{purple}{rgb}{0.65, 0.12, 0.82}
 
\lstdefinelanguage{JavaScript}{
keywords={typeof, new, true, false, catch, function, return, null, catch, switch, var, if, in, while, do, else, case, break},
keywordstyle=\color{blue}\bfseries,
ndkeywords={class, export, boolean, throw, implements, import, this},
ndkeywordstyle=\color{darkgray}\bfseries,
identifierstyle=\color{black},
sensitive=false,
comment=[l]{//},
morecomment=[s]{/*}{*/},
commentstyle=\color{purple}\ttfamily,
stringstyle=\color{red}\ttfamily,
morestring=[b]',
morestring=[b]"
}
 
\title[IGM-Vis]{IGM-Vis: Analyzing Intergalactic and Circumgalactic Medium Absorption Using Quasar Sightlines in a Cosmic Web Context}
\author[J. N. Burchett, D. Abramov, J. Otto, C. Artanegara, J. X. Prochaska \& A. G. Forbes]{\parbox{\textwidth}{\centering J. N. Burchett$^{1}$, D. Abramov$^{2}$, J. Otto$^{2}$, C. Artanegara$^{2}$, J. X. Prochaska$^{1}$, and A. G. Forbes$^{2}$ }
        \\
{\parbox{\textwidth}{\centering $^1$Department of Astronomy and Astrophysics, University of California, Santa Cruz, United States\\
         $^2$Department of Computational Media, University of California, Santa Cruz, United States
       } 
}
}

\begin{document}\teaser{
 \includegraphics[width=\linewidth]{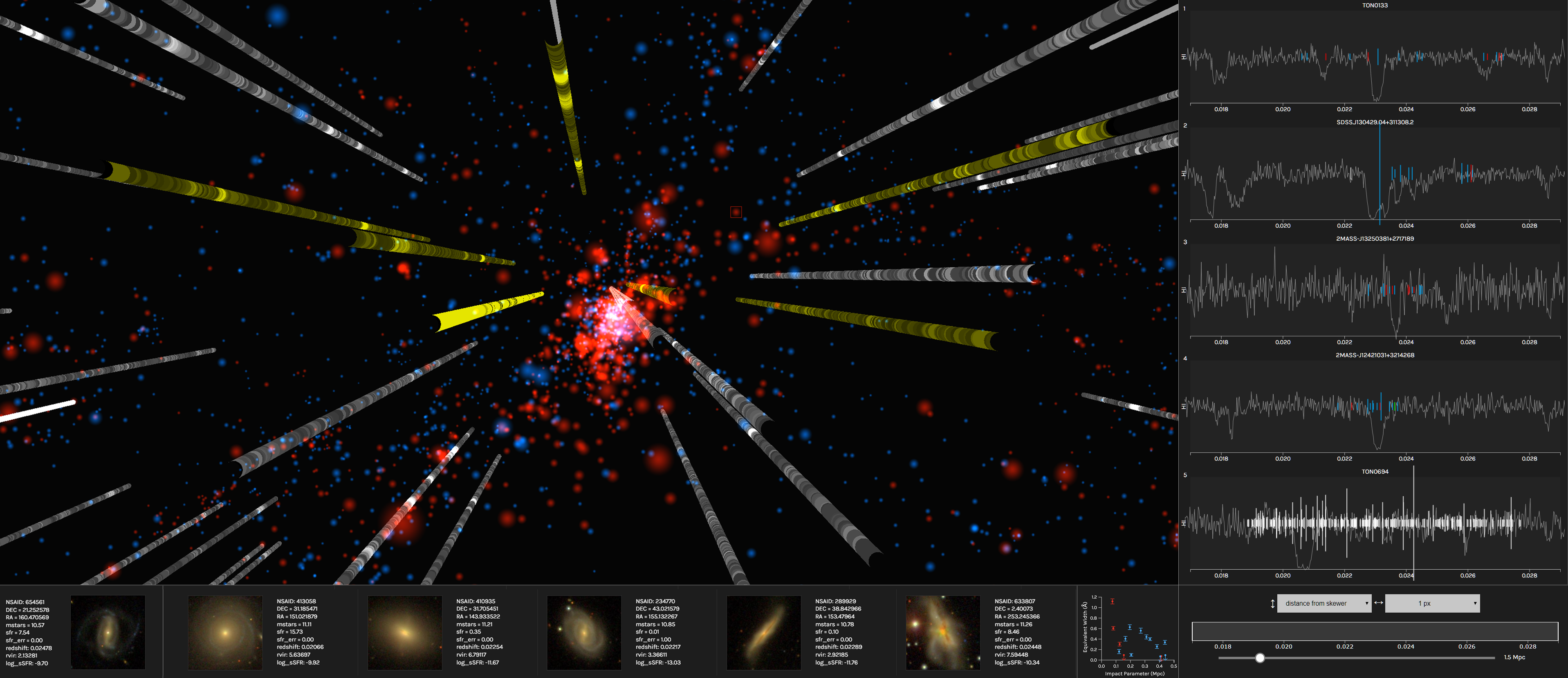}
 \centering
  \caption{A screen capture of the \textit{IGM-Vis} visualization application, which facilitates a range of analysis tasks using quasar sightline data in order to better understand intergalactic medium and circumgalactic medium absorption features. The Universe Panel (upper left) shows a 3D map of galaxies in the Coma Supercluster, along with ``skewers'' representing absorption signals in spectra of background quasar sightlines. The Galaxy Panel (lower left) provides descriptive metrics for selected galaxies. The Spectrum Panel (right) displays the spectra and marks nearby galaxies, facilitating comparative analysis between galaxies and absorption, and provides interactive controls with to select which regions of the 3D map are visible, to choose the zoom level of the spectra, and to update the profile plot to its left in the Equivalent Width Plot Panel. }
\label{fig:teaser}
}

\maketitle
%-------------------------------------------------------------------------

\begin{abstract}We introduce \textit{IGM-Vis}, a novel astrophysics visualization and data analysis application for investigating galaxies and the gas that surrounds them in context with their larger scale environment, the Cosmic Web.  Environment is an important factor in the evolution of galaxies from actively forming stars to quiescent states with little, if any, discernible star formation activity. The gaseous halos of galaxies (the circumgalactic medium, or CGM) play a critical role in their evolution, because the gas necessary to fuel star formation and any gas expelled from widely observed galactic winds must encounter this interface region between galaxies and the intergalactic medium (IGM). We present a taxonomy of tasks typically employed in IGM/CGM studies informed by a survey of astrophysicists at various career levels, and demonstrate how these tasks are facilitated via the use of our visualization software. Finally, we evaluate the effectiveness of \textit{IGM-Vis} through two in-depth use cases that depict real-world analysis sessions that use IGM/CGM data.




\end{abstract}

%-------------------------------------------------------------------------
\section{Introduction}

Since the first astrophysical discoveries of absorption lines in the spectra of distant quasars over 50 years ago \cite{Schmidt:1966aa}, the lines-of-sight to these objects that skewer the vast, intervening expanses of the Universe have served as indispensable probes of cosmic structure formation and evolution~\cite{Bahcall:1969lr,Bahcall:1969qy}.  These absorption signatures reveal complexes of gas intersected by the sightline; these gas complexes, which compose the intergalactic medium (IGM), inhabit the relatively dense pockets of the cosmic void that also contain galaxies, galaxy groups, and galaxy clusters.  On their own, IGM spectra enable scientists to study the evolution of `metal' (in the parlance of astrophysics, any element heavier than Helium) enrichment in the Universe, the conditions immediately after the Big Bang, and the thermodynamic history of the Universe.  However, connecting the gas complexes detected in quasi-stellar object (QSO, a class that includes quasars) spectra to particular structures requires substantial ancillary data, which in turn require their own investment of valuable telescope resources, in order to place the IGM spectra in context.  In particular, galaxy surveys around the QSO sightlines enable researchers to associate particular absorption features with galactic environments, such as galaxy halos (the circumgalactic medium, or CGM) or the intermediary gas within galaxy clusters.

Galaxies show incredible diversity in their shapes, sizes, and colors.  However, this diverse landscape partitions broadly into two categories: those that are actively forming stars and those that are not forming stars at appreciable rates~\cite{Baldry:2004qy}. Young, massive stars are bluer in color than long-lived, lower-mass stars, which appear red in color; thus star-forming (SF) galaxies appear bluer than `red and dead' quiescent galaxies, which have few young stars and primarily contain older populations.  Understanding how galaxies evolve from one state to another has become a cornerstone of astrophysics.  Galaxies do not evolve as `closed boxes', as evidenced by their abundances of heavy elements~\cite{Tinsley:1974zl}, the fact that galaxies cannot sustain star formation without being fed by gas from their surroundings~\cite{Larson:1980vn}, and the gaseous outflows they expel~\cite{Veilleux:2005lr}.  This dynamic view of galaxy evolution has brought the CGM into sharp focus, as this interface region between galaxies and the IGM likely hosts many of the processes involved~\cite{tumlinson2017circumgalactic}.

Since the launch of the Hubble Space Telescope (HST) in 1990 and especially the installation of the Cosmic Origins Spectrograph (COS) \cite{Green:2012qy} instrument in 2009, the astrophysical community has amassed large quantities of both the QSO spectra and accompanying galaxy surveys to conduct these analyses, but the analysis methods employed to date have largely been limited to imposing set criteria during sample selection and focusing the ensuing analysis on searching for correlations within those limited parameters. Punzo et al.~\cite{punzo2015role} advocate for the use of visualization tools to assist astrophysicists with a richer set of analysis tasks, taking note of the common pitfalls in many of the current applications used by the astrophysics community, which include the complexity of user interfaces and the lack of interactive analysis features. %, and the inability to scale effectively to larger datasets.

We present \textit{IGM-Vis}, a novel visual analytics software application that facilitates more sophisticated IGM and CGM analysis tasks than are available in existing analysis tools and that provides a series of integrated 2D and 3D views of galaxies, quasar sightlines, and analysis plots. The development of \textit{IGM-Vis} was informed by conversations with dozens of astrophysicists who work with these datasets on a daily basis as part of their research, including in-depth interviews with eight IGM/CGM experts who spent a significant amount of time with the application. In addition to the contribution of \textit{IGM-Vis}, we introduce a taxonomy of tasks relevant to the \mbox{IGM/CGM} community, derived from a comprehensive survey of astrophysicists at different career levels, and provide detailed use cases illustrating the scientific workflow of astrophysicists, and that highlight the effective use of \textit{IGM-Vis} for IGM/CGM identification, analysis, and presentation tasks. Astrophysics terminology that may be less familiar to visualization researchers is defined throughout the paper, and summarized in Table~\ref{tab:SummaryOfTerms}.

%-------------------------------------------------------------------------
\section{Background \& Related Works}
\label{Sec:RelatedWork}

A state-of-the-art review of observational and theoretical CGM research is presented by Tumlinson et al.~\cite{tumlinson2017circumgalactic}, which emphasizes the importance of the CGM within the larger context of galaxy evolution. The CGM can be roughly defined as the gaseous envelope surrounding a galaxy, with a size often expressed as the galaxy's \textit{virial radius}, the approximate maximum distance for which matter is gravitationally bound.  Gas flows between the IGM, the CGM, and the interstellar medium, and characteristics of the gas are typically observed by measuring absorption lines in the spectra of light emitting objects behind the gas clouds. Visualizations used in contemporary astrophysics research include spectral plots, which show the data directly and reveal the absorption from material along sightlines, and absorption profile plots such as \textit{equivalent width}, which measures the absorption line strength, versus the projected distances of nearby galaxies. \textit{IGM-Vis} generates interactive versions of these plots on-the-fly for selected quasar spectra, making it easy to quickly associate galaxies with their imprints upon the absorption spectra.

The landmark COS-Halos survey~\cite{tumlinson2013cos} investigates the CGM of forty-four galaxies by selecting both star forming (SF) and quiescent galaxies over a range of mass. Key results of this survey are that the CGM exhibits strong absorption of neutral hydrogen (\hone) for both quiescent and SF galaxies \cite{Thom:2012lr}, and that the CGM contains at least half of all the non-dark matter in galaxy. Other studies also find that galaxies are correlated with the strongest \hone~absorbers, with the weaker absorbers likely tracing diffuse cosmic filaments and the IGM~\cite{Chen:2005kx,Prochaska:2011yq}. \textit{IGM-Vis} provides a novel interface for analyzing both IGM and CGM data, and enables researchers to investigate the relationships among galaxies, cosmic structure, and absorption patterns.

Cosmological simulations based on the cold dark matter paradigm predict that matter in the Universe is organized into a Cosmic Web (also known as large-scale structure), as elongated, interconnected filaments formed from dark matter contain low density IGM gas as well as galaxies and their CGM~\cite{Rauch:1998aa}. Indeed, most of the non-dark matter mass in the Universe likely resides in the IGM \cite{Cen:1999yq}.  In regions of the Universe nearer to our own Milky Way, large surveys can reveal the Cosmic Web traced by galaxies~\cite{Geller:1989aa}. A study by Wakker et al.~\cite{wakker2015nearby} uses HST/COS to probe one Cosmic Web filament and its imprint of H I absorption lines. \textit{IGM-Vis} facilitates the analysis of multiple filament structures using quasar sightline data. 

A range of visualization tools have been created to mitigate the complexity of astrophysics data. Popular web applications, such as the The Sloan Digital Sky Survey's SkyServer \cite{York:2000aa} and the World Wide Telescope \cite{gray2004world,rosenfield2018aas}, compile and present an enormous amount of astronomical image data. The European Space Agency's \textit{ESASky}~\cite{baines2016visualization} provides access to data from multiple astronomical archives, and can display the sky at different wavelengths. However, these websites do not provide any tools to analyze the data directly. Similarly, \textit{mViewer}~\cite{berriman2017application} enables a user to merge multiple image layers, using an image mosaic engine to project multiple 2D images into common astronomical layouts. \textit{AstroShelf}~\cite{neophytou2012astroshelf} also facilitates querying multiple datasets, enabling a scalable navigation of data and data annotations. Sagrist{\`a} et al.~\cite{sagrista2018gaia} introduce visualization tools to navigate observations made by the Gaia Spacecraft. Luciani et al.~\cite{luciani2014large} introduce an interface to control the transparency of multiple image layers so that relevant data from multiple datasets can be seen at the same time. Work by Boussejra et al.~\cite{Boussejra2018} leverages visual programming techniques to filter and analyze multi-spectral datasets. \textit{IGM-Vis} emphasizes the presentation and analysis of spectrum data, and contextualizes these spectra with images for user-selected regions of the Universe on demand.

A number of tools present astrophysical elements as volumes within a 3D view~\cite{fu2007transparently,taylor2017visualizing}. For example, Pomar{\`e}de et al.~\cite{pomarede2017cosmography} make use of images, videos, and derived isosurface structures within a 3D representation to show galaxy position, velocity and density fields, gravitational potential, and velocity shear tensors. Punzo et al.~\cite{punzo2015role} also note the importance of coupling 3D views with alternative visual representations, and emphasize interactive data filtering in order to investigate relevant elements. Popov et al.~\cite{popov2012analyzing} explore methods to visualize dynamical structures in cosmological simulation data, showing how 3D plots can be used to compare the resulting outputs from various computational methods. Haroz et al.~\cite{haroz2008multiple} include a 2D parallel coordinates plot alongside a 3D visualization to emphasize uncertainty inherent to an astronomical dataset or when found through a comparison of datasets. Fujishiro et al. introduce \textit{TimeTubes}~\cite{fujishiro2018timetubes}, which transforms temporal blazar data into an unusual volumetric structure, using ellipses to encode polarization parameters arranged as a 3D ``tube'' in order to identify patterns of interest. \textit{IGM-Vis} represents galaxies as an interactive 3D scatterplot in which particular regions of the Universe are pierced by cylindrical representations of sightlines, which can then be more thoroughly examined via linked 2D spectral plots.

Visual analytics tools have been used to explore simulation data that model the evolution of the Universe~\cite{hanula2015cavern,preston2016integrated}. Almryde and Forbes~\cite{almryde2015halos} introduce an interactive web application to visualize ``traces'' of dark matter halos as they move in relation to each other over time, and Scherzinger et al.~\cite{scherzinger2017interactive} present an innovative application that provides 2D and 3D views to support the analysis of halo substructures and hierarchies. \textit{IGM-Vis} also provides a visual analytics dashboard comprised of integrated panels~\cite{Dang2017_BioLinker_IEEEPacificVis,ForbesFontana_VAST2018,Ma2015_JIST_SwordPlots,Sarikaya_dashboard2018}, facilitating a workflow supporting IGM/CGM identification, analysis, and presentation tasks. In addition to these standalone software applications, a range of frameworks and platforms have been developed to support astrophysical visualization. This includes work by Woodring et al.~\cite{woodring2011analyzing}, which uses ParaView~\cite{ahrens2005paraview,ayachit2015paraview} to analyze cosmological simulation data, and the Aladin Sky Atlas ~\cite{bonnarel2000aladin,boch2014aladin}, which enables users to add annotation markers to image data catalogs. Tools such as TOPCAT and STIL~\cite{taylor2005topcat} and Glue~\cite{beaumont2015hackable} are useful for generating and exploring tabular datasets and to explore relationships within and across related datasets. \textit{IGM-Vis} focuses specifically on facilitating analyses of quasar sightlines and their nearby galaxies.

%------------------------------------------------------------------------

\begin{table}

\begin{tabular}{ l|l } 
\multicolumn{1}{l|}{\textbf{Data Tasks}} & 
\multicolumn{1}{l}{\textbf{Description}} 
 \\
\begin{tabular}{@{}l@{}}\textbf{T1}: Obtain Sightline\\Spectra\end{tabular} 
&

\begin{tabular}
{@{}l@{}}Query archives; Make telescope \\ observations \end{tabular}

\\ \hline 

\begin{tabular}
{@{}l@{}}\textbf{T2}: Obtain Galaxy \\Data \end{tabular}
&
\begin{tabular}
{@{}l@{}}Derive measurements from \\ spectroscopy and imaging \end{tabular}
\\ \hline \hline

\multicolumn{1}{l|}{\textbf{Identification Tasks}} & 
\multicolumn{1}{l}{\textbf{Description}}  \\
\begin{tabular}
{@{}l@{}}\textbf{T3}: Identify \\Foreground Features\end{tabular} 
&
\begin{tabular}
{@{}l@{}} Identify galaxies near sightlines; \\ Identify larger structures
\end{tabular}
\\ \hline 

\begin{tabular}
{@{}l@{}}\textbf{T4}: Measure \\Absorption Properties
\end{tabular}
&
\begin{tabular}
{@{}l@{}}Find absorption associated with \\ galaxies or structures \end{tabular}
\\ \hline 

\begin{tabular}
{@{}l@{}}\textbf{T5}: Identify Sightline \\ Features
\end{tabular}
& 
\begin{tabular}
{@{}l@{}} Find relevant features across \\ multiple sightlines \end{tabular}

\\ \hline \hline
\multicolumn{1}{l|}{\textbf{Analysis Tasks}} & 
\multicolumn{1}{l}{\textbf{Description}}  \\
\begin{tabular}
{@{}l@{}}\textbf{T6}: Test Correlations
\end{tabular}
& 
\begin{tabular}
{@{}l@{}} Quantify relationship between \\absorption and galaxies
\end{tabular}
\\ \hline 

\begin{tabular}
{@{}l@{}}\textbf{T7}: Discover \\Absorption Patterns
\end{tabular}
& 
\begin{tabular}
{@{}l@{}} Compare multiple sightlines; \\Generate hypotheses from \\ analyzing sightlines
\end{tabular}
\\ \hline \hline

\multicolumn{1}{l|}{\textbf{Presentation Tasks}} & 
\multicolumn{1}{l}{\textbf{Description}}  \\
\begin{tabular}
{@{}l@{}}\textbf{T8}: Create Derived \\Datasets
\end{tabular}
&
\begin{tabular}
{@{}l@{}} Share data with astrophysics \\community 
\end{tabular}
\\ \hline 

\begin{tabular}
{@{}l@{}}\textbf{T9}: Produce Plots
\end{tabular}
& \begin{tabular}
{@{}l@{}} Create plots for presentations;\\ Explore results interactively 
\end{tabular}
\\ \hline

\end{tabular}

\caption{A taxonomy of tasks relevant to the IGM/CGM community.} \label{tab:Tasks}
\end{table}

\begin{figure}
 \includegraphics[width=3.3in]{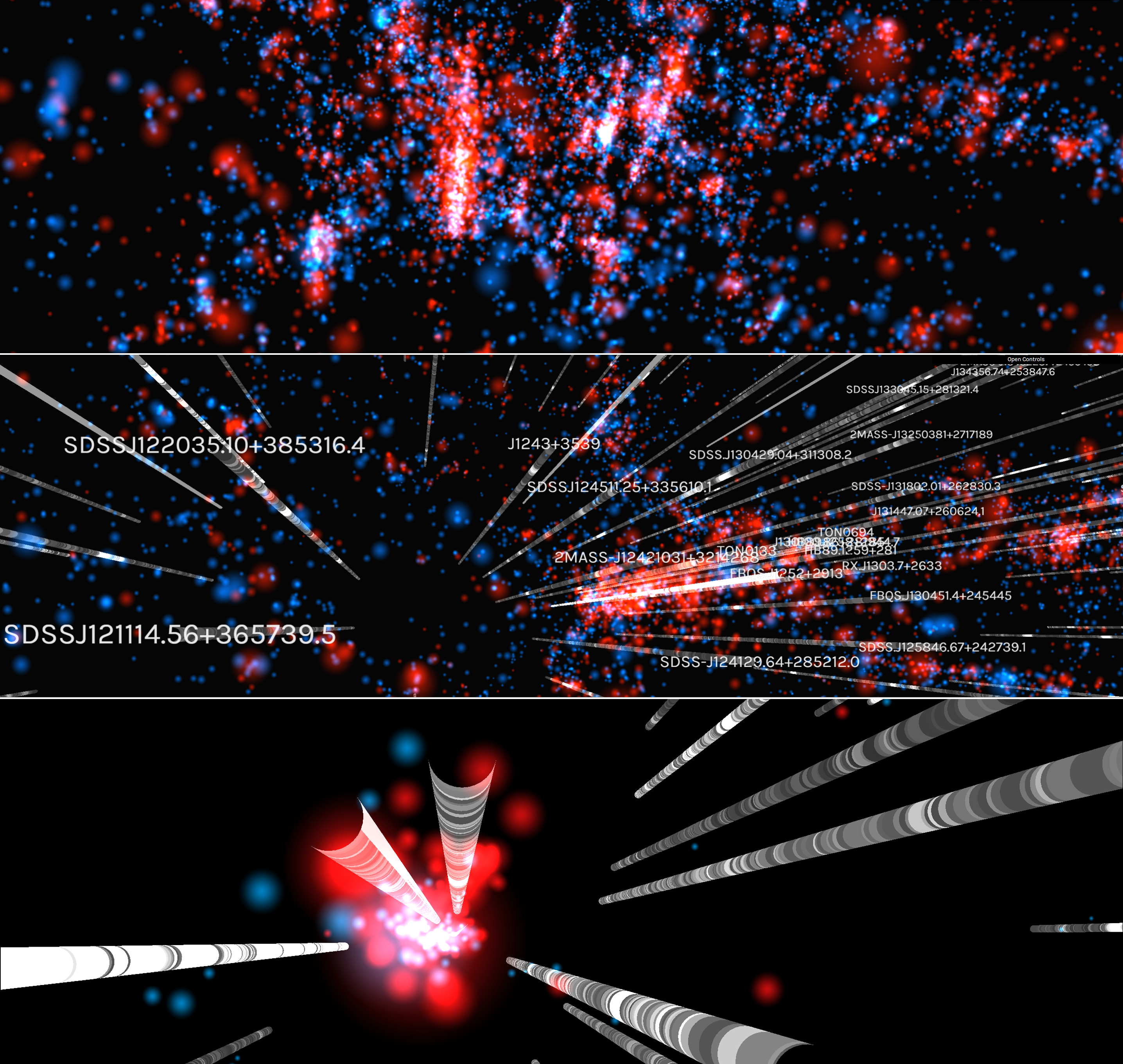}
 \centering
  \caption{Zooming into a specific region of the Coma Supercluster dataset in the Universe Panel. To reduce visual clutter, a user can toggle on or off different elements or filter the number of galaxies displayed. The top image displays an overview of a large number of galaxies; the middle panel zooms into a region of interest, with skewers and labels toggled on; the bottom image filters out galaxies beyond a user-specified distance threshold from the skewers.}
\label{fig:galaxyTryp}
\end{figure}

\section{IGM/CGM Task Analysis}
\label{Sec:AnalysisTasks}

Although astrophysicists utilize a diverse set of very large datasets in their research, including data from ground and space telescopes, data transformed by computational models, and simulation data, to date there are no best practices for effective scientific workflows investigating IGM/CGM data, nor a comprehensive overview of primary analysis tasks relevant to astrophysicists using sightline data. Lud{\"a}scher et al.~\cite{ludascher2006scientific} present a formalized system to define scientific workflows, and McPhillips et al.~\cite{mcphillips2009scientific} and Etemadpour et al.~\cite{etemadpour2015designing} present practical approaches to implementing steps within a scientific workflow. Task analyses are a useful way to determine how visualization tools can effectively support cognition ~\cite{pirolli2005sensemaking,tory2004human}. Isenberg et al.~\cite{isenberg2008grounded} advocate for evaluating visualization tools situated within the context of their intended use, and Lam et al.~\cite{lam2012empirical} introduce scenario-based approaches to evaluating visualization tools, which emphasize understanding environments and work practices. In order to determine the ``why'' and ``how'' of abstract visualization tasks~\cite{brehmer2013multi}, we conducted a structured task analysis involving both observations of astrophysics labs and surveys of astrophysics researchers at different career levels. An initial survey was conducted with 40 astrophysicists engaged in analyzing QSO data to determine the visualization needs of the community, and we conducted further in-depth interviews with 8 IGM/CGM experts in order to characterize the specific tasks they perform.

All survey respondents focus primarily on extragalactic astrophysical research (studying objects beyond our own Milky Way), and all are primarily observational astronomers who analyze data from a variety of sources.  Each actively collects data from various telescope facilities but also relies heavily on publicly available archives, including the Sloan Digital Sky Survey (SDSS) and the Hubble Space Telescope (HST) archive, which are the two datasets we analyze with \textit{IGM-Vis}. The data products from these sources include not only imaging and spectroscopy that our researchers analyze by, e.g., measuring absorption lines in quasar spectra, but also compiled catalogs of tabular data, such as galaxy positions and brightnesses measured by the SDSS team. 

Our respondents indicated challenges due to the lack of shared or distributed data and software, and they also noted that visualization components are not sufficiently integrated within their current software tools. Based on our observations and conversations with researchers, responses from our survey, and a review of recent literature, we defined a list of primary tasks involved in IGM/CGM analysis, which we categorize as \textit{data collection} tasks, \textit{identification} tasks, \textit{analysis} tasks, and \textit{presentation} tasks (summarized in Table~\ref{tab:Tasks}). In addition to guiding the development of \textit{IGM-Vis}, we expect that this taxonomy will be of use for other visualization researchers who plan to create tools for astrophysicists and for students and astrophysicists from other fields who wish to understand the primary activities of \mbox{IGM/CGM} researchers.

\subsection{Data Collection Tasks}

Studies of the IGM/CGM using absorption line spectroscopy primarily require the spectra of QSOs or galaxies, against which the foreground gas will absorb. These spectra are acquired via new observations or from archival sources, such the SDSS~\cite{York:2000aa}, Hubble Spectroscopic Legacy Archive~\cite{Peeples:2017aa}, or Keck Observatory Archive \cite{OMeara:2015aa}.  Because different telescopes and instruments are sensitive to different wavelengths or have other differing characteristics (e.g., spectral resolution), spectra from multiple sources for a single sightline will often be compiled to provide diagnostics from multiple spectral lines. These spectra often require normalization before taking measurements of absorption lines. 

The defining characteristic of CGM studies relative to those focusing on the IGM is the knowledge or supposition of galaxies whose halos are probed by the sightline spectra.  Relating the CGM absorption characteristics to the properties of host galaxies requires associating the two by their \textit{redshifts}. Redshift is a measure of velocity, calculated from the observed wavelengths of intrinsic spectral features that have been shifted to longer (redder) wavelengths.  The redshift provides the best estimate of distance for objects far outside our own Galaxy, as Hubble's Law~\cite{Hubble:1929aa} and cosmological models provide the link between distance and redshift, which is measured most precisely by spectroscopy.  The default dataset we study in \textit{IGM-Vis} includes only foreground galaxies for which spectra exist, and our ensuing workflow description assumes datasets that include spectroscopic information about these galaxies in addition to spectra of the sightlines probing them.

\noindent \textit{\textbf{T1: Obtain spectra of objects probing foreground environment}}
This task may involve taking new telescope observations or querying one or more archival sources when different wavelength ranges are covered by different telescopes/instruments.

Redshifts are typically measured from the locations of known spectral features, such as strong emission or absorption lines. A researcher typically will correlate CGM absorption properties with properties of the `host' galaxies, which must be derived from the galaxy spectra and/or imaging.  These include a galaxy's luminosity, color (ratio of flux between two photometric bands), mass in stars (stellar mass, $M_*$), dark matter halo mass ($M_{halo}$), star formation rate (SFR), and metallicity.  Several methods exist for measuring these quantities from both imaging and spectroscopy, including directly using spectral line fluxes in calibrated formulae \cite{Kewley:2002lr,Kennicutt:2012qy} and fitting the spectral energy distributions (SEDs) with stellar, nebular, and dust emission models \cite{Walcher:2011aa}. 

\noindent \textit{\textbf{T2: Obtain information on foreground objects composing the probed environment}}
This task involves deriving measurements from spectroscopy and/or imaging, unless relevant catalogs of derived measurements are already available.

\subsection{Identification Tasks} 

Astrophysical datasets often contain billions of data points, and the datasets are growing: as of 2015, astronomical instruments produce twenty-five zettabytes of data annually~\cite{stephens2015big}.  To feasibly define scientific questions and the samples of data to address them, one must select objects and attributes of interest from the troves of information. \textit{IGM-Vis} integrates sky survey data, which provide information and images of individual galaxies, with quasar sightline data, which provide information about the nature of gas in different regions of space. A primary research task is to search for and identify data within a certain region of the Universe that is associated with user-specified features. 

A purely IGM study may not require any information about foreground galaxies; the sample to analyze may be selected by absorption features identified with a particular ion, such as \hone\ \cite{Wolfe:1986aa,Lehner:2007lr}, \osix\ \cite{Tripp:2000aa,Lehner:2014kq}, \cfour \cite{Songaila:2001qy,Burchett:2015rf}, or a combination of these.  CGM studies may follow either what are sometimes known as `galaxy-selected' or `absorption-selected' approaches.  In the former, samples comprise particular galaxies with desired properties according to selection criteria, and their sightline probes are analyzed for absorption at similar redshifts.  In the latter, one attempts to associate absorption features in background spectra with foreground galaxies projected around the sightline.

Recent CGM studies consider relationships to galaxy environment and focus on objects in dense clusters \cite{Burchett:2018aa,Yoon:2013kq} or groups \cite{Burchett:2015aa,Pointon:2017aa}.  These environments as well as filaments and voids may be investigated from a more galaxy-agnostic perspective to reveal the gas composition, kinematics, and physical conditions in dense and underdense regions of the Universe \cite{Tejos:2012lr,Tejos:2016qv,wakker2015nearby}. \textit{IGM-Vis} is ideally suited for selecting targets based on environment, as the user can quickly identify filaments and dense structures visually and leverage the data available to generate hypotheses and test them on the fly. 

\noindent \textit{\textbf{T3: Identify features in foreground environment to investigate with spectral probes}}
This task involves identifying galaxies nearby to sightlines and identifying larger structures comprising galaxies, such as clusters, filaments, or voids.

If adopting a galaxy- or environment-selected strategy, the task then turns to measuring the associated absorption in spectra probing the foreground objects.  Often, a researcher adopts some velocity window near the redshift of interest (such as that of the host galaxy) that determines the wavelength range in the spectrum where absorption due to some species transition (such as \hone\ \lya) will be associated to the foreground object. \textit{IGM-Vis} includes the ability to quickly measure absorption for hypothesis testing and exploratory analysis, and assists the researcher in identifying absorption features that can be more extensively analyzed offline.

\noindent \textit{\textbf{T4: Measure absorption properties associated with foreground structures}}
This task involves identifying absorption properties near the redshifts of foreground structures and searching for coherent absorption associated with structures across multiple sightlines.

A researcher may also conduct tomography of foreground structures wherein multiple sightlines pierce the halo of a single foreground galaxy~\cite{Bowen:2016aa}, galaxy cluster~\cite{Yoon:2017aa}, or filament~\cite{wakker2015nearby}.  Alternatively, structures may be identified through absorption alone, as multiple adjacent sightlines might exhibit coherent absorption at a consistent redshift~\cite{Cai:2016aa,Lee:2018aa}. \textit{IGM-Vis} enables researchers to visually select multiple sightlines within a single filament or galaxy cluster, and encourages the interactive investigation of their absorption properties.

\noindent \textit{\textbf{T5: Identify features in multiple spectra to investigate origin}}
This task involves identifying interesting or coherent features across several sightlines.

\begin{figure*}
 \includegraphics[width=\textwidth]{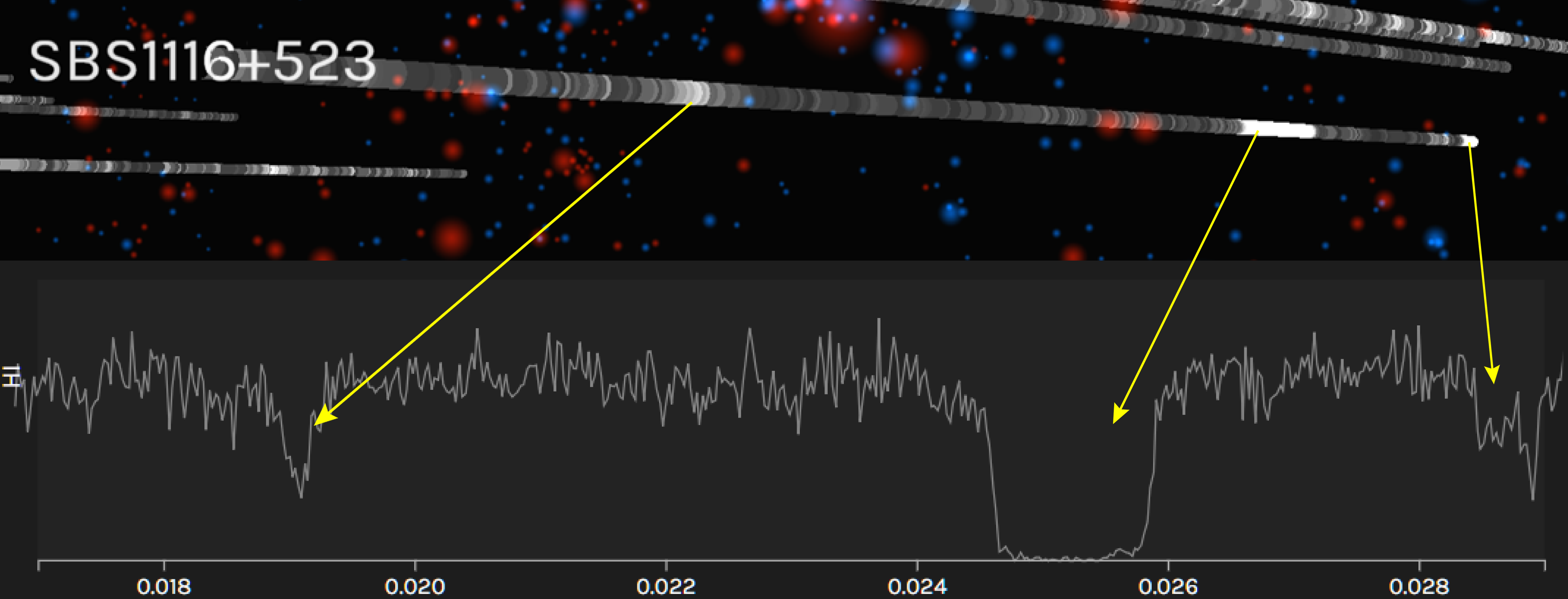}
 \centering
  \caption{The same ``skewer,'' sightline SBS1116+523, represented in the Universe Panel (top) and in the Spectrum Panel (bottom). On the top, the brighter coloring along the skewer indicates absorption, which a user can explore in more detail in the associated spectral plot. }
\label{fig:spectra3d}
\end{figure*}

\subsection{Analysis Tasks}

Once galaxies and sightlines of interest have been identified, researchers analyze them in terms of a variety of metrics, depending on their specific goals or hypotheses. Insight into the formation and evolution of galaxies emerges from observable trends or shared characteristics among the galaxy populations. Perhaps the most important result of a CGM survey is the absorption profile, or absorption strength as a function of impact parameter.  If the foreground galaxy redshifts are known, then the \textit{impact parameter} (the projected physical distance between a galaxy and a sightline) can be calculated. The absorption strength is typically quantified by the equivalent width and/or column density, with \textit{equivalent width} being readily measured from any normalized spectrum.  The equivalent width is simply defined as the width of a rectangle with area equal to that between the absorption line and normalized flux level ($=1$), if the height of the rectangle extended from zero flux to 1.  Hence, the equivalent width-impact parameter relation is often the first (and often key) result of any CGM study. Much valuable information is encoded in these absorption profiles: How far does the CGM extend?  How does the absorption strength change with increasing distance from the galaxy center?  How does the profile change as measured in different ions?  How do the profiles of galaxies in more isolated environments compare to those in dense structures? The answers to these questions have strong implications for galaxy formation models, and \textit{IGM-Vis} enables researchers to reason about these models effectively. 

\noindent \textit{\textbf{T6: Test correlations between absorption and foreground objects}} This task involves investigating galaxy and/or environmental properties and absorption metrics, such as by quantifying the relationship between equivalent width and impact parameter.

Any observational dataset is, to some degree, limited by the signal-to-noise ratio (S/N).  At any S/N, weak enough absorption features will be `buried in the noise', rendering statistically insignificant detections even if present.  Also, archival datasets compiled from heterogeneous sources may comprise observations from programs with vastly different data quality requirements.  Thus, in IGM spectra, one sightline might have a vastly greater S/N than another.  To address both of these cases, one can employ stacking to boost the S/N in a composite spectrum.  In such analysis, collections of sightlines are chosen based on some criterion (e.g., those that probe star forming galaxies within 100 kpc), the spectra are all transformed to some reference frame (e.g., the redshifts of galaxies they probe), and the spectra are coadded by, e.g., the average or median flux value at each pixel~\cite{Borthakur:2015aa}.  

\noindent \textit{\textbf{T7: Discover weak absorption by stacking data from multiple sightlines}} This task involves generating and testing hypotheses about galaxies, clusters, and filaments through a process that merges spectra of subpar sensitivity or from heterogeneous datasets in order to isolate and characterize absorption features not observable in individual spectra.

\subsection{Presentation Tasks}

The standard for sharing large data tables and images in the astrophysics community is the Flexible Image Transport System (FITS) format. For IGM work, data tables commonly include information about the QSOs observed and measurements of absorption systems. CGM studies often include this information as well as the properties of associated galaxies, such as stellar mass and SFR. Researchers can use \textit{IGM-Vis} to export these data for galaxy-sightline pairs of interest.

\noindent \textit{\textbf{T8: Build and release IGM/CGM datasets}} This task involves creating, exporting, and publishing data tables containing derived measurements from analyzing galaxies and sightlines, so that other researchers can utilize results for future studies.

Current interactive visualization tools for astrophysical data do not accommodate the distinct aspects of \mbox{CGM/IGM} research. \mbox{CGM/IGM} science results are typically presented using static plots. For observational studies, these often include `stack plots' of the spectral data showcasing one or more several transitions within the same absorption system transformed to a uniform velocity frame of reference.  The relationships among various absorption and galaxy properties in CGM studies are generally reflected in 2D scatterplots~\cite{Werk:2013qy} and may include theoretical model predictions when the comparable model data are available and/or relevant. \textit{IGM-Vis} features the use of interactive plots that facilitate data sharing, enabling other researchers to validate analyses or to provide information that supports or challenges a hypothesis. 

\noindent \textit{\textbf{T9: Produce static or interactive plots}} This task involves creating static plots for presenting data in scientific articles and presentations or to communicate with the public, and it may involve producing interactive plots that can be used to summarize results, to illustrate novel ideas, to annotate interesting features, and to validate hypotheses.

%------------------------------------------------------------------------

\section{The \textit{IGM-Vis} Application}
\label{Sec:IGM-Vis}
In this section, we provide an overview of \textit{IGM-Vis}, and discuss how our design decisions promote the analysis tasks described in Section~\ref{Sec:AnalysisTasks}. (Although IGM and CGM research are both enabled by \textit{IGM-Vis}, we chose to title the application \textit{IGM-Vis} as IGM datasets underlie the work in both fields.) By default, \textit{IGM-Vis} provides coverage of the Coma Supercluster and its surroundings to the extent covered by the SDSS. The dataset we employ consists of QSO spectroscopy from the Hubble Spectroscopic Legacy Archive~\cite{Peeples:2017aa} and galaxy information provided by the NASA/Sloan Atlas (NSA)~\cite{Blanton:2011lr,sdssManga}, supporting \textbf{T1} and \textbf{T2}. Additionally, using the NSA galaxy catalog, we calculate a measurement of star formation rate,  enabling \textbf{T3}. Two key preprocessing steps are conducted prior to integrating the QSO data into \textit{IGM-Vis}: the QSO spectra are trimmed to include only the wavelength range where our two spectral diagnostics (\hone\ and \cfour) would be observed over the redshift range of interest, and we normalize the spectra by fitted continua to enable spectral measurements directly within IGM-Vis, supporting \textbf{T4}. For the Coma Supercluster data, we used a subset of astrophysical data localizing on galaxies and quasars that fall within a right ascension (RA) range of $115^{\circ}$ and $260^{\circ}$, a declination (DEC) range between $-4^{\circ}$ and $65^{\circ}$, and a redshift (z) range between 0.018 and 0.023. This resulted in 19,268 galaxies and 348 quasar sightlines containing  \hone~and/or \cfour\ absorption data. Though we designed \textit{IGM-Vis} around this initial dataset~\cite{2017hstprop15009B}, many other quasar spectral lines, wavelength ranges, and redshift regions of the Universe can be visualized within \textit{IGM-Vis}. 

A core design decision in developing \textit{IGM-Vis} is to enable and encourage users to begin their analysis from various starting points and to take different paths during an investigation of the Cosmic Web. To that end, \textit{IGM-Vis} is a modular platform composed of four primary panels, each of which provides a different view of astrophysical data: (1) an interactive 3D visualization of galaxies and QSO sightlines, or ``skewers''; (2) image data and metadata from the SDSS for selected galaxies; (3) interactive 2D plots of spectra for selected skewers; and (4) an equivalent width profile plot that is generated dynamically by user interaction. Fig.~\ref{fig:teaser} shows an overview of the application. \textit{IGM-Vis} enables comparisons between absorption features in a single QSO sightline and its surrounding galaxies as well as comparisons between multiple QSOs simultaneously. This is useful for identifying absorption patterns of a spectral line that may be related to particular features of neighboring galaxies. One key use is to quickly visually identify cosmic filaments~\cite{wakker2015nearby} and inspect the influence these structures may have on their gas.

\subsection{Universe Panel}
The main panel provides an interactive 3D plot of the angular position and distance of all galaxies and quasar sightlines in the dataset, supporting the identification tasks \textbf{T3} and \textbf{T5}. Including a 3D plot was an important design feature requested by the majority of \mbox{IGM/CGM} researchers we interviewed during the development of the project, as it provides a Cosmic Web context to make it easier to reason about relationships between sightlines and galaxies within particular redshift ranges, which can be challenging when using 2D scatterplots representations. Galaxies are represented as partly transparent colored circles (Gaussian blurs), where blue represents star-forming galaxies and red represents quiescent galaxies, and where the size of the galaxy indicates its virial radius. Sightlines are represented as cylindrical ``skewers'' and colored differently along their length to indicate the amount of absorption in the spectrum (by default, neutral hydrogen \hone~absorption), where dark grey indicates no absorption and white (or yellow when selected) indicates strong absorption. Regions of strong \hone\ absorption appear as bright bands on the skewer cylinders, and a user can quickly discern which galaxies reside near high-absorption regions. The skewers and galaxies are all rendered over a black background, and skewers are outlined in yellow when they are selected by a user. Both galaxy color maps and skewer color maps can be customized by the user via a drop-down options menu. Fig.~\ref{fig:galaxyTryp} shows different views of the Universe Panel at different zoom levels and with interactive filtering applied.

Each galaxy and skewer is positioned according to their angular coordinates in the celestial sphere: right ascension (RA) and declination (DEC). Navigating the 3D view is controlled using keyboard shortcuts or via the mouse. Text displaying the name of each skewer and the visibility of the skewers themselves can be toggled on or off using either the drop-down menu or a keyboard shortcut.

Several computations are performed on the data in order to be effectively presented in the application. As astrophysical objects are measured in projection on the sky, object redshifts are used in transformations into physical distance. Each galaxy and data point along a skewer has a corresponding redshift, which are converted to physical distances (units of Megaparsecs, or Mpc) via cosmological formulae and plotted in 3D space. We then convert from spherical coordinates by using the RA and DEC angles, lookup the corresponding physical distance for each redshift, and output a 3D position vector.

\begin{figure}
 \includegraphics[width=3.3in]{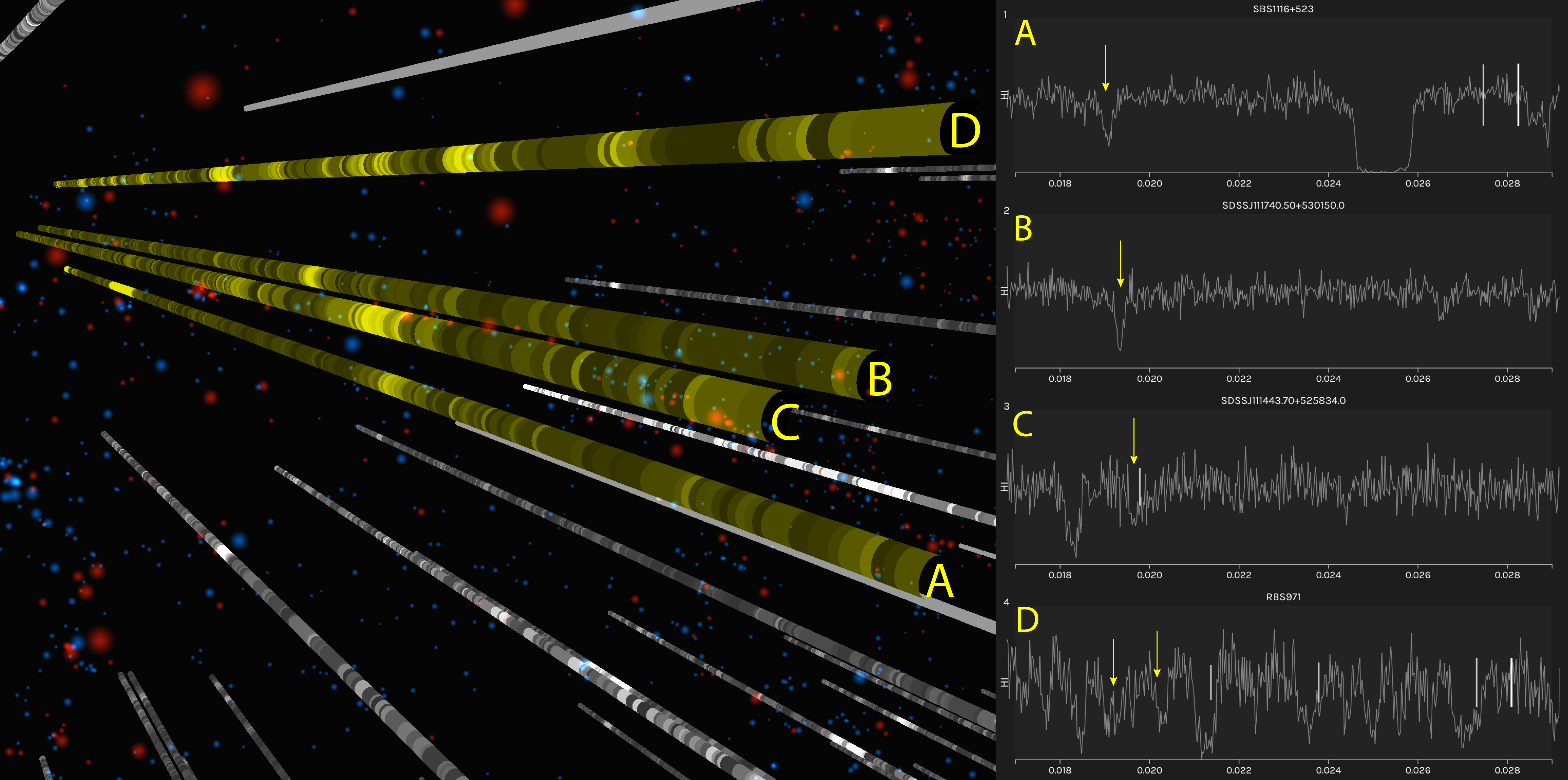}
 \centering
  \caption{A screen capture of the \textit{IGM-Vis} interface while investigating Use Case 1. 
  The Universe Panel shows 4 skewers with absorption (brighter shading) at similar redshifts, and
  the Spectrum Panel shows each sightline's spectrum with the absorption features of interest labeled
  with yellow arrows.  The slider below these spectra (not shown here) has been set to mark the redshifts of galaxies
  within $\sim$500 kpc of the skewers (white vertical lines), only the SDSSJ111443.70+525834.0 sightline has a galaxy 
  present near the redshift of interest (green vertical line). }
\label{fig:4sightlines}
\end{figure}

\subsection{Galaxy Panel}
Directly below the Universe Panel, information about selected galaxies is displayed along the bottom of the application window and is updated each time a user hovers over a galaxy in the Universe Panel. Each galaxy contains a list of attributes: its unique identifier (NSAID), DEC, RA, stellar mass (mstars), star formation rate (sfr), star formation rate uncertainty (sfrerr), a log of the specific star formation rate (log\_sSFR), redshift, and the virial radius (rvir). When one hovers over a galaxy, this information is displayed along with its corresponding image, retrieved from the SDSS~\cite{Abolfathi:2018aa}. A user can interactively select and store galaxies of interest, which will then continue to populate the Galaxy Panel even after the user has moved the mouse off of that galaxy. These stored galaxies are also highlighted in the Spectrum Panel, as we discuss below, using either a blue or red tick mark to show the galaxy's redshift within the spectral plots if it's impact parameter is within a user-selected threshold of the currently selected skewers. The bottom of Fig.~\ref{fig:teaser} shows a diverse set of galaxies found within the Coma Supercluster dataset. The Galaxy Panel  provides context in support of the identification tasks (especially \textbf{T3}), and relevant galaxy data can be exported for further analysis (\textbf{T8}).

\begin{table*}

\begin{center}
\begin{tabular}{ l|c|c|c|c|c|c|c|c|c } 
\multicolumn{1}{l|}{\textbf{Dataset}}  & 
 \multicolumn{1}{l|}{\textbf{Galaxies}} &
 \multicolumn{1}{l|}{\textbf{Skewers}} &
 \multicolumn{1}{l|}{\textbf{Redshift Range}} &
 \multicolumn{1}{l}{\textbf{Data Size}} &
 \multicolumn{1}{l}{\textbf{Speed Score}} &
 \multicolumn{1}{l}{\textbf{TTI}} &
 \multicolumn{1}{l}{\textbf{IL}} &
  \multicolumn{1}{l}{\textbf{FPS MBP}} &
 \multicolumn{1}{l}{\textbf{FPS PC}} 
 
 \\\hline 
Small & 4160 & 41 & [0.0168, 0.0298] & 36.4 MB & 98/82 & 1.2s/5.3s & 10ms/10ms & $\sim$60fps & $\geq$60fps \\ \hline 

Medium & 19268 & 348 & [0.0168, 0.0298] & 217.8 MB & 98/80 & 1.2s/5.6s & 10ms/40ms & $\sim$36fps & $\sim$57fps \\ \hline 

Large & 37663 & 348 & [0.0128, 0.0348] & 257.4 MB & 98/78 & 1.3s/5.8s & 10ms/60ms & $\sim$26fps & $\sim$49fps \\ \hline

\end{tabular}
\end{center}
\caption{Performance metrics for \emph{small}, \emph{medium}, and \emph{large} IGM/CGM datasets. The \emph{small} dataset includes galaxies and sightlines in the Coma Supercluster, while the \emph{medium} and \emph{large} datasets include additional galaxies surrounding the Coma Supercluster. In addition to the number of galaxies, skewers, and the redshift range, we include metrics from Google's PageSpeed Insights (https://developers.google.com/speed/), including the overall Speed Score, the time-to-interactive (TTI) metric, and the input latency (IL) for both desktop and mobile application simulations, along with the average frames per second (FPS) for a) a 15-inch 2012 model MacBook Pro laptop with a 2.6 GHz Intel Core i7-3720qm CPU and an NVIDIA GeForce GT 650M CPU running macOS 10.14, and b) a custom-built PC with a 4.00 GHz Intel Core i7-8086k CPU and an NVIDIA Titan XP Pascal 12 GB GPU running Windows 10.
} \label{tab:Performance}
\end{table*}

\subsection{Spectrum Panel}
The Spectrum Panel is located on the right side of the application, primarily supporting analysis tasks \textbf{T4} and \textbf{T6}. When a skewer is hovered over in the Universe Panel, it appears in the topmost position of the Spectrum Panel. Fig.~\ref{fig:spectra3d} shows an example of how bright spots on ``skewers'' in the Universe panel are represented in the Spectrum Panel. Multiple spectral plots can be stored within the panel, making it easy to compare absorption profiles, and each skewer can contain multiple spectral plots (\hone~and \cfour\ in the Coma Supercluster dataset). The x-axis of each plot is in units of redshift, and the y-axis represents normalized flux. 

The range of redshift values displayed can be filtered using an interactive slider, which is mapped to all of the spectral plots for consistent comparison. Colored tick marks are used to represent galaxies within a user-specified impact parameter. The user can interactively toggle the visibility of galaxies beyond this distance within the Universe Panel. The relative height and width of these tick marks can be interactively mapped to different attributes in the galaxy data, such as its impact parameter, virial radius, stellar mass, or star formation rate. Fig.~\ref{fig:4sightlines} illustrates a coordinated analysis using the Universe Panel and Spectrum Panel, described in Use Case 1. The user can also export a file that contains all data within the Spectrum Panel, including the name and spectra for each skewer, along with a list of all nearby galaxies within a specified impact parameter, supporting tasks \textbf{T8} and \textbf{T9}.

\subsection{Equivalent Width Plot Panel}
\begin{figure*}
 \includegraphics[width=\textwidth]{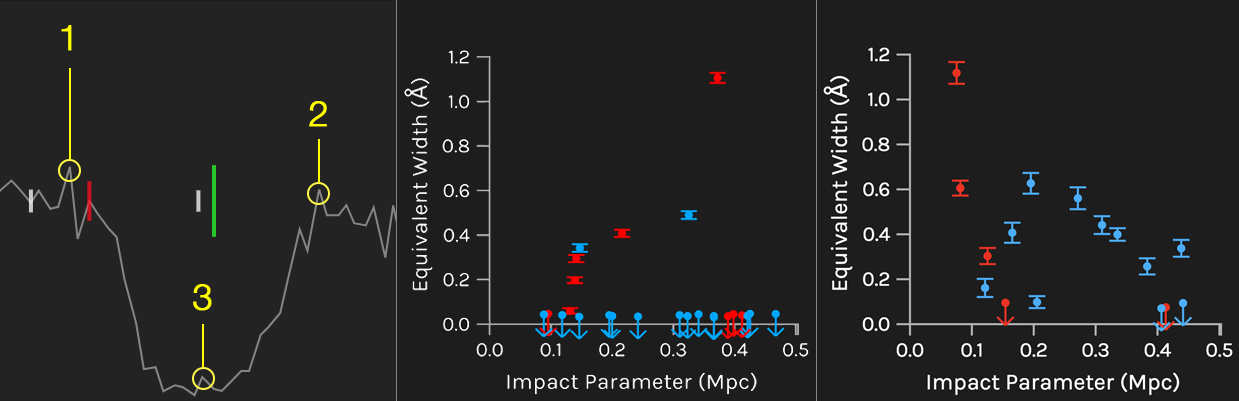}
 \centering
  \caption{The far left panel shows an example of how an equivalent width measurement is captured: the user selects (1) the left boundary, (2) the right edge, and (3) a center reference point. The two equivalent width profiles shown in the middle and right panels correspond to Use Case 2. The middle plot shows CGM measurements within filaments near the Coma Cluster, resulting in mostly nondetections, whereas the rightmost plot shows measurements taken in regions $>50$ Mpc away from the Coma Cluster, resulting in a greater incidence of detections.}
\label{fig:EquivalentPlot3}
\end{figure*}

Positioned between the Galaxy Panel on the bottom and the Spectrum Panel on the right is a plot for visualizing the projected distance of a quasar sightline-galaxy pair (impact parameter, x-axis) and the absorption strength (equivalent width, y-axis) of a user selected spectral region. This plot is dynamically generated on demand once an equivalent width measurement is made by selecting boundary points on a spectral line. \textit{IGM-Vis} calculates the equivalent width of the spectral feature and plots the value against the impact parameter of the nearest galaxy to the spectral skewer. The data points are dynamically filtered to within a user-specified impact parameter range and/or a user-specified redshift range.  This panel supports identification task \textbf{T4} and analysis tasks \textbf{T6} and \textbf{T7}, and these plots can be exported for inclusion in presentations (supporting \textbf{T9}). Fig.~\ref{fig:EquivalentPlot3} illustrates how an equivalent width plot is interactively populated from within the Spectrum Panel, along with two example outputs created for Use Case 2.

\subsection{Implementation Details}
\label{SubSec:Impl}

\textit{IGM-Vis} runs on any modern web browser, and makes use of \textit{three.js} for 3D visualization in the Universe Panel and \textit{D3.js} for data management and plotting the interactive 2D graphs and displaying galaxy metadata. Data is preprocessed prior to being uploaded to the application using custom software functions made available with the software that: a) translates the cosmological dataset into Cartesian coordinates used in the Universe Panel, and b) creates a lookup table to enable realtime filtering of galaxies by either redshift or by distance from selected galaxies or sightlines. Fraedrich et al.~\cite{fraedrich2009exploring} and Schatz et al.~\cite{schatz2016interactive} introduce rendering techniques to display very large datasets that may include trillions of points. However, for our initial use cases--- which focus on a relatively narrow region of the Universe, as is common for IGM/CGM analysis--- we ingest smaller datasets consisting of between 4,160 and 37,663 galaxies and between 41 and 348 skewers. Running on a desktop computer, \textit{IGM-Vis} renders up to 20,000 galaxies and 400 skewers at close to realtime rates, while a larger dataset encompassing a wider redshift range runs at well above interactive rates. (See Table~\ref{tab:Performance} for details.)

\begin{table*}

\begin{tabular}{ l|c|c|c } 
\multicolumn{1}{l|}{\textbf{Title}}  & 
 \multicolumn{1}{l|}{\textbf{Research Areas}} &
 \multicolumn{1}{l|}{\textbf{Datasets}} &
 \multicolumn{1}{l}{\textbf{Software Tools}} 
 \\\hline
Professor & IGM, CGM, FRB & \begin{tabular}{@{}c@{}}Keck HIRES, KCWI, \\ HST, CASBaH\end{tabular} & linetools, pyigm \\ \hline 

Assistant Professor & CGM, Galaxy spectroscopy & \begin{tabular}{@{}c@{}}HST COS, HRC, Keck HIRES, \\  Gemini GMOS, HST WFC3\end{tabular} & \begin{tabular}{@{}c@{}}linetools, pyigm, veeper,\\ redrock, specdb, IDL\end{tabular} \\ \hline 

Assistant Professor & 
\begin{tabular}{@{}c@{}}CGM, Galactic winds, \\Quasar absorption line systems\end{tabular} &
\begin{tabular}{@{}c@{}}Galaxy and quasar \\ spectroscopy, redshift catalogs\end{tabular} &
\begin{tabular}{@{}c@{}}linetools, specdb, \\ sdss-marvin, pyigm\end{tabular} 
\\ \hline 

Assistant Professor & 
\begin{tabular}{@{}c@{}}CGM, Galaxy evolution, \\Galaxy halo-gas connection\end{tabular} &
Spectroscopy, imaging databases & 
custom python modules 
\\ \hline

Postdoctoral Researcher & 
\begin{tabular}{@{}c@{}}Extra-galactic astronomy, Galactic \\outflows, Cosmic reionization\end{tabular} &
\begin{tabular}{@{}c@{}}The Hubble Legacy Archive, \\SDSS, GALEX\end{tabular} &
custom IDL programs
\\ \hline 

PhD Student & 
\begin{tabular}{@{}c@{}}FRB, Spectra analysis of galaxies\end{tabular} &
\begin{tabular}{@{}c@{}}SDSS, Spectroscopy data\end{tabular} &
python
\\ \hline 

PhD Student & 
\begin{tabular}{@{}c@{}}Dwarf galaxies, \\Primordial helium abundance\end{tabular} &
\begin{tabular}{@{}c@{}}SDSS\end{tabular} &
\begin{tabular}{@{}c@{}}CasJobs, PypeIt, emcee, \\custom python modules\end{tabular} 
\\ \hline 

PhD Student & 
CGM &
\begin{tabular}{@{}c@{}}CGM$^2$, COS, Gemini GMOS\end{tabular}  &
\begin{tabular}{@{}c@{}}pyigm, guesses\end{tabular}  
\\ \hline 

\end{tabular}
\caption{Title and research areas of the experts who provided feedback, along with the datasets and software tools they commonly use. } \label{tab:Experts}
\end{table*}

\section{Evaluation}
\label{Sec:Eval}
We invited eight experts (four male and four female) at different career stages--- a subset of those surveyed regarding workflow and analysis tasks whose responses are summarized in Section~\ref{Sec:AnalysisTasks}--- to spend time with \textit{IGM-Vis} and to provide feedback on their experiences. Four are faculty at research institutions, three are graduate students, and one is a postdoctoral researcher. All employ spectroscopy of galaxies in their research, which includes investigations of fast radio bursts (FRB), galactic winds, and dwarf galaxies, along with IGM and CGM data analyses. Although none of them uses the Coma Supercluster dataset provided for their evaluation in their own work, they were able to navigate the data without any issues, and indicated their familiarity with the spectral analysis and equivalent width profile plots. Table~\ref{tab:Experts} lists relevant details about these experts, including their career level, research areas, and which software tools they commonly use for IGM/CGM analysis.

The reaction from our survey respondents was overwhelmingly positive, with each of the experts noting the novelty of using interactive visualization software for identification and analysis tasks, and described \textit{IGM-Vis} using terms such as ``fantastic,'' ``great,''  and ``impressive.'' One respondent noted that \textit{IGM-Vis} was useful for exploratory analysis, that it provided an intuitive way to ``get a feel for the data'', and appreciated that no special installation or downloads were necessary to run the software. Another noted that it gave her ``an appreciation for data visualization,'' as similar visualization tools do not exist for analyzing IGM/CGM data. The experts reported a diverse set of potential applications for \textit{IGM-Vis}, with multiple users suggesting that identifying interesting configurations of galaxies, sightlines, and absorbers would be aided by \textit{IGM-Vis}. One respondent highlighted the ease of interaction and noted how straightforward it was to compare different galaxies and sightlines. Another wrote that \textit{IGM-Vis} is ``a powerful tool to diagnose which galaxies correspond to which intervening absorption systems.'' Yet another concluded that \textit{IGM-Vis} ``is extremely well-suited to diagnose the physical mechanism that leads to CGM absorption.'' Our experts also lauded the potential for scientific outreach given the difficulty of describing astrophysics research methods to the general public. One respondent told us that \textit{IGM-Vis} could be ``useful for both experts and non-experts in the field,'' and appreciated the ability to share the data and visualization easily in order to facilitate reproducibility. Another researcher described an insight gleaned from experimenting with \textit{IGM-Vis}, expressing surprise that several skewers showed absorption features while no galaxies existed were nearby within 1 \rvir\ and, conversely, was surprised that there were a number of galaxies at small impact parameter but did not have detectable absorption, despite the high covering fraction of \hone\ in the CGM~\cite{Prochaska:2011yq,Stocke:2013mz}. Two of the experts explicitly described new hypotheses generated while using \textit{IGM-Vis} and provided details about the process of developing new ideas and performing  initial investigations involving star formation rates and filament features. 

\subsection{Use Case 1: Investigating Coherence of Multisightline Absorption Signals}
\label{ss:uc1}

While the large scale structure traced out by galaxies is quite conspicuous in the Universe
panel, absorption features are also apparent as bright regions on the skewers. This enables
one to quickly explore the absorption features that may by associated with individual 
galaxies or identify absorption features that appear in multiple sightlines at similar 
redshifts, both within the context of the large scale
environment. Here, we investigate the nature of a coherent multi-sightline 
absorption signal discovered using \textit{IGM-Vis}, providing an example of \textbf{T5} (investigating origins) and \textbf{T6} (testing correlations).

Using multiple sightlines to reveal the spatial structure of absorbing media by sampling several points across the plane of the sky has been employed on a variety of scales. This tomographic approach has been applied to the gas clouds in our own Milky Way~\cite{Tripp:2012lr,Fox:2014lr,Bordoloi:2017ab}, 
the halos of other galaxies~\cite{Bowen:2016aa,Zahedy:2016aa,Rubin:2018ab,Lopez:2018aa}, and, on the largest scales, the intergalactic medium of the Cosmic Web~\cite{Cai:2016aa,Lee:2018aa}. Within the \textit{IGM-Vis} volume, we identified at least four sightlines that exhibit \hone\ absorption
signals within a narrow redshift range of one another.  Fig.~\ref{fig:4sightlines} shows
an interface view with these skewers visible in the Universe Panel along with the four spectra of interest
shown in the Spectrum Panel.  We have annotated the absorption features at z$\sim$0.019 in
these panels.  Remarkably, these sightlines are separated
by as much as $\sim$2.5 Mpc, and only one sightline passes within 500 kpc of a detected 
galaxy having a similar redshift as the absorption.  Using the 3D
navigation in the Universe Panel, we find that a putative filamentary structure of 
galaxies passes through this group of sightlines at slightly higher redshift (z$\sim$0.021).
Although potentially related to these galaxies, the absorbing complex we have discovered would 
have velocity separations of $>300$ km/s from the approximate central redshift of the group
of galaxies near the sightline skewers (crudely estimated from the data presented in the
Galaxy Panel of IGM-Vis).  If not bound to the same underlying dark matter infrastructure, 
the gas may be separated by $\gtrsim 5$ Mpc.

\lya~absorption is nearly ubiquitous within galaxy halos, extending to their virial radii 
and at least to 3 \rvir \cite{Tripp:1998kq,Prochaska:2011yq, Johnson:2015xy,Burchett:2015aa}. Weak 
absorption, extending to larger distances may actually trace the filaments and Cosmic Web 
structures hosting galaxies themselves, and \cite{Prochaska:2011yq} has estimated the
sizes of the weak-\lya\  traced filaments to be $\sim $400kpc.  Thus, if the coherent absorption
we detect across these four sightlines probes the same gas complex, the size of the
system would far exceed this scale and does not coincide with the nearby structure traced by 
galaxies.

\subsection{Use Case 2: Investigating Absorption Patterns among Galaxy Filaments using Equivalent Width Plots}
\label{ss:uc2}

The Cosmic Web is composed of vast filaments, sheets, and walls traced by galaxies held together by a skeleton of dark matter.  The intersections of these filaments, nodes, are where massive clusters of galaxies form, and these sites of galaxy cluster formation are the densest pockets of the Universe.  Here, we explore the hypothesis that the CGM of galaxies within filaments that are in closer proximity to the massive Coma cluster are preferentially stripped of their gas relative to those in apparently less dense filaments further away.

We begin by using the slider below the Spectrum Panel to mark the redshifts of galaxies within $\sim$500 kpc of the sightline skewers.  We selected skewers that: (a) probed putative filament structures near Coma, and (b) had galaxies with impact parameters  $\lesssim 500$ kpc.  We then measured the equivalent widths at the redshifts marked in each spectrum panel.  As shown in the equivalent width profile panel in Fig.~\ref{fig:EquivalentPlot3}, we registered mostly nondetections at all impact parameters.  This is a bit surprising, because we selected galaxies well outside Coma itself (but within filaments \textit{near} Coma), and a high detection rate of \hone\ is expected \cite{Burchett:2018aa}. Next, we selected skewers that also probed filaments but those that are (a) well separated from Coma on the sky, translating to distances of $>50$ Mpc and (b) had galaxies with impact parameters  $\lesssim 500$ kpc.  Similarly, we measured equivalent widths at the redshifts of these galaxies.  Intriguingly, we measured 12 \hone\ detections, with only 4 nondetections at all impact parameters.  

Our experiment within \textit{IGM-Vis} appears to validate our hypothesis: the higher detection rate of \hone\ and the large \lya\ equivalent widths we measure for the non-proximate filaments to Coma are consistent with a picture where the CGM in filaments `feeding' Coma are indeed preferentially stripped.  We acknowledge the dangers of possible confirmation bias in selecting the skewers in each category, but a unique hallmark of \textit{IGM-Vis} is the ability test and retest such hypotheses \textit{in mere minutes}.  From here, we will proceed with a more rigorous analysis of the data to establish the statistical significance of this result.  As we have identified several skewers with nearby galaxies that do not exhibit detectable absorption, we can easily extract this list of sightlines from \textit{IGM-Vis} (\textbf{T8}) and stack the spectra to check whether a signal emerges that is too weak for detection in the native S/N of the data (supporting \textbf{T7}).

\section{Conclusion \& Future Work}

\textit{IGM-Vis} was developed through an iterative design process that included multiple rounds of feedback both from astrophysicists and visualization researchers over a 13 month period between February 2018 and March 2019. The use cases presented above demonstrate that \textit{IGM-Vis} is already empowering astrophysical investigation from a wholly different, environmental context-sensitive perspective than those commonly employed by the IGM/CGM community. In particular, \textit{IGM-Vis} enables a range of identification, analysis, and presentation tasks that are not well supported by other visualization tools.  Based on the feedback from astronomers with varying interests and at different stages of their careers, we have additionally identified a range of data collection and preprocessing tasks that are not currently supported directly in \textit{IGM-Vis}. Indeed, currently, the the most challenging aspect of our application is transforming heterogeneous data sources into a form that can be ingested into \textit{IGM-Vis}. However, we have developed open source JavaScript functions that simplify astrophysics data processing; for future work, we plan to incorporate them directly into the visualization. These tools are freely available via our GitHub project repository at \url{https://github.com/CreativeCodingLab/Intergalactic}, along with detailed instructions on how to use \textit{IGM-Vis} and import custom data, an interactive web demo that was used for Use Case 1 and Use Case 2 and to create all figures in this paper, documented source code, and a video tutorial.

\begin{table*}
\centering
\begin{tabular}{l|l}

 \multicolumn{1}{l|}{\textbf{Term}} &
 \multicolumn{1}{l}{\textbf{Definition}} 
 
 \\\hline

 IGM, intergalactic medium  &\begin{tabular}{@{}l@{}}The diffuse gas between galaxies. Often defined as material  found beyond the \rvir\ of galaxies\end{tabular}  \\ \hline 
 CGM, circumgalactic medium & \hspace{1.6mm} Gaseous halos surrounding galaxies. Often defined as material within \rvir\   \\ \hline 
\rvir\, virial radius & \begin{tabular}{@{}l@{}} A fiducial radius from the center of a galaxy to which the CGM extends. Within this radius,\\  gas is considered likely to be gravitationally bound to the galaxy\end{tabular} \\ \hline 
Equivalent width & \begin{tabular}{@{}l@{}}A measure of absorption strength for a spectral line that indicates the amount of material along \\ the line of sight\end{tabular} \\ \hline 
QSO, Quasi-stellar object & \begin{tabular}{@{}l@{}}Quasar or other object used as background source for absorption line spectroscopy. Skewers \\ in IGM-Vis are generated from QSO spectra \end{tabular} \\ \hline
HST, Hubble Space Telescope & \begin{tabular}{@{}l@{}}HST is an ultraviolet/optical/infrared observatory launched in 1990, and is the source \\ of all QSO spectra for IGM-Vis' default dataset\end{tabular}  \\ \hline
COS, Cosmic Origins Spectrograph & \hspace{1.6mm} COS is the instrument aboard HST that obtained QSO spectra for IGM-Vis' default dataset  \\ \hline
SFR, star formation rate & \begin{tabular}{@{}l@{}}The star formation rate is typically expressed in units of the mass of the Sun per year. Galaxies \\ are colored in IGM-Vis according to their SFR relative to their mass \end{tabular}  \\ \hline
SDSS, Sloan Digital Sky Survey & \begin{tabular}{@{}l@{}}SDSS is a large imaging and spectroscopic survey that provides all galaxy information \\ within IGM-Vis' default dataset\end{tabular} \\ \hline
$z$, redshift & \begin{tabular}{@{}l@{}}Redshift is a measure of velocity calculated from the observed wavelengths of spectral \\ features that have been shifted to longer (redder) wavelengths. It is used to estimate and \\ represent  distance for galaxies in IGM-Vis\end{tabular} \\ \hline
RA, right ascension & \hspace{1.6mm} A celestial coordinate that, along with DEC, defines a position on the sky  \\ \hline
DEC, declination & \hspace{1.6mm} A celestial coordinate that, along with RA, defines a position on the sky \\ \hline
H~\textsc{i}, neutral hydrogen  & \hspace{1.6mm} Primary tracer of IGM and CGM gas within IGM-Vis' default dataset \\ \hline
C~\textsc{iv}, triply ionized carbon  & \hspace{1.6mm} Tracer of ionized, heavy element-enriched CGM and IGM gas within IGM-Vis' default dataset \\ \hline
Ly$\alpha$, Lyman $\alpha$ &  \begin{tabular}{@{}l@{}}Spectral transition from the ground state to the first excited state of H~\textsc{i}. Ly$\alpha$ is a spectral \\ feature  indicated in IGM-Vis' skewers and measured for equivalent width profiles \end{tabular}  \\ \hline
kpc, kiloparsec & \hspace{1.6mm} An astrophysical unit of distance, approximately 3,260 light years or $3 \times 10^{16}$ km \\ \hline
Mpc, megaparsec & \hspace{1.6mm} An astrophysical unit of distance, equal to 1000 kpc  \\
\hline
\end{tabular}
\caption{Definitions of astrophysics terms and abbreviations used throughout this article.}
\label{tab:SummaryOfTerms}

\end{table*}

\section{Acknowledgements}
This research is funded by Hubble Space Telescope grant \mbox{HST-AR \#15009}. Hardware used to develop the project is provided through the Nvidia GPU Grant program.

%-------------------------------------------------------------------------

\bibliographystyle{eg-alpha-doi}

%\bibliography{IGM-Viz-2019}
\newcommand{\etalchar}[1]{$^{#1}$}

\end{document}